\documentclass[pra, aps, longbibliography, twocolumn, reprint, superscriptaddress, showpacs, footnoteinbib, 
floatfix]{revtex4-1}
\usepackage[english]{babel}
\usepackage[latin1]{inputenc}
\usepackage[dvipsnames]{xcolor}
\usepackage{graphicx}
\usepackage{amssymb}   % for math
\usepackage{amsfonts}
\usepackage{amsmath}
\usepackage{pdfpages}
\usepackage{dcolumn}   % needed for some tables
\usepackage{bm}        % for math
\usepackage[mathscr]{eucal}
\usepackage[colorlinks=true, linkcolor=OrangeRed,citecolor=RoyalBlue,urlcolor=NavyBlue]{hyperref}
\usepackage[all]{hypcap} % let hyperlinks correctly point to figures rather than their captions; 
\usepackage[caption=false]{subfig}

\newcommand{\be}{\begin{equation}}
\newcommand{\ee}{\end{equation}}
\newcommand{\bea}{\begin{eqnarray}}
\newcommand{\eea}{\end{eqnarray}}

\newcommand{\bK}{{\bf K}}

\newcommand{\ci}{\mathfrak{i}}

\definecolor{darkred}{rgb}{0.6,0,0}
\definecolor{darkblue}{rgb}{0.0,0,0.6}
\definecolor{red}{rgb}{1,0,0}

\newcommand{\cA}{c_\text{A}}
\newcommand{\cB}{c_\text{B}}
\newcommand{\calC}{{\cal C}}
\newcommand{\calG}{{\cal G}}

\newcommand{\nzm}{n_\text{zm}}
\newcommand{\NA}{{N_\text{A}}}
\newcommand{\NB}{{N_\text{B}}}

\newcommand{\Nzm}{{N_\text{zm}}}
\newcommand{\Ndb}{{N_\text{db}}}
\newcommand{\nAB}{n_\text{vac}}

\newcommand{\zp}{\zeta_\text{p}}
\newcommand{\szm}{SZM} 
\newcommand{\ds}{site}

\begin{document} 
% \preprint{p10.zeromodes-DoS}

\title{Graphene with vacancies: supernumerary zero modes}

\author{Norman Weik}
\email{Current affiliation: Institute of Transport Science, RWTH Aachen University, D-52056 Aachen, Germany}
\affiliation{ Institute of Nanotechnology,
 Karlsruhe Institute of Technology, Campus North, D-76344
  Karlsruhe, Germany}
\affiliation{Institut f\"ur Theorie der Kondensierten Materie,
 Karlsruhe Institute of Technology, Campus South, D-76128 Karlsruhe, Germany}
\author{Johannes Schindler}
\email{Current affiliation: Institut f\"ur Technische Optik, Pfaffenwaldring 9, 70569 Stuttgart.}
\affiliation{ Institute of Nanotechnology, Karlsruhe Institute of Technology, Campus North, D-76344 Karlsruhe, Germany}
\affiliation{Institut f\"ur Theorie der Kondensierten Materie, Karlsruhe Institute of Technology, Campus South, D-76128 Karlsruhe, Germany}
%\author{Victor H\"afner}
%\email{Current affiliation:  Institute for Information Management in Engineering (IMI),  Karlsruhe Institute of
%Technology, 76131 Karlsruhe.}
%\affiliation{ Institute of Nanotechnology,
% Karlsruhe Institute of Technology, Campus North, D-76344
% Karlsruhe, Germany}
% \affiliation{Institut f\"ur Theorie der Kondensierten Materie,
% Karlsruhe Institute of Technology, Campus South, D-76128 Karlsruhe, Germany}
\author{Soumya Bera}
\affiliation{Max-Planck-Institut f\"ur Physik komplexer Systeme, 01187 Dresden, Germany}
\author{Gemma C. Solomon}
\affiliation{Nano-Science Center, University of Copenhagen, DK-2100 Copenhagen, Denmark}
\affiliation{Department of Chemistry, University of Copenhagen, DK-2100 Copenhagen, Denmark}
\author{Ferdinand Evers}
\affiliation{ Institute of Theoretical Physics,
 University of Regensburg, D-93050 Regensburg, Germany}

\date{\today}% It is always \today, today,

%\keywords{ }%Use showkeys class option if keyword
\begin{abstract}

The density of states, $\varrho(E)$, of graphene is 
investigated within the tight binding (H\"uckel) approximation 
in the presence of vacancies. 
%($\NA$, $\NB$) 
They introduce a non-vanishing density of zero modes, 
$\nzm$, that act as midgap states: $\varrho(E)=n_\text{zm}\delta(E) + \text{smooth}$. 
As is well known, the actual number of zero modes per sample can, in
principle, exceed the sublattice imbalance: 
$N_\text{zm}\geq |\NA-\NB|$, where $\NA$, $\NB$ denote the number of 
carbon atoms in each sublattice. 
In this work, we establish a stronger relation that is valid in the 
thermodynamic limit and that  involves the {\it concentration}
of zero-modes: 
$n_\text{zm}>|\cA-\cB|$, where $\cA$ and $\cB$ denote the 
concentration of vacancies per sublattice; in particular, 
$n_\text{zm}$ is non-vanishing even in the case 
of balanced disorder, $\NA/\NB=1$. 
Adopting terminology from benzoid graph theory, the excess modes associated with the current carrying backbone (percolation cluster) are called {\it supernumerary}. 
In the simplest cases, such modes can be associated 
with structural elements such as carbon atoms connected with a single bond, only. 
Our result suggests 
that the continuum limit of bipartite hopping models \ 
supports nontrivial ``supernumerary'' terms that escape the present continuum descriptions.  
\end{abstract}
%\pacs{PACS numbers: }
%\narrowtext
\maketitle
%%%%%%%%%%%%%%%%%%%%%%%%%%%%%%%%%%%%%%%%%%%%%%%%%%%%%%%%%%%%%%%%%%%%%%5
\section{Introduction}
%%%%%%%%%%%%%%%%%%%%%%%%%%%%%%%%%%%%%%%%%%%%%%%%%%%%%%%%%%%%%%%%%%%%%%5
% {\it Introduction.}
As is well known, Dirac-particles are realized, e.g., 
in clean graphene, which exhibits two Dirac-cones at the 
$\bK$ and $\bK'$ points of the Brillouin zone. 
\cite{CN09}
Another important incarnation is met in energy-spectra 
of quasi-particles over (two dimensional, 2D) 
condensates of fermions with $p-$type pairing.~\cite{read00,volovik}
% Such condensates can be realized, e.g., 
% in $5/2$-quantum Hall liquids or $p-$type superconductors. 
In three dimension, the Dirac dispersion 
can be realized in Weyl- and Dirac-type 
semimetals.~\cite{Weyl11,Hosur13}

%%%%%%%%%%%%%%%%%%%%%%%%%%%%
One, out of many interesting aspects of Dirac-fermions 
relates to their topological properties.~\cite{CN09,SAR10}
When punching a hole (``defect'') into Dirac-gases 
% of these qp 
% -- e.g. by local suppression of the order parameter -- 
a zero-energy state can form at its boundary. 
% Due to fermion-doubling these modes represent paired 
% half-Dirac fermions , i.e. {\em Majorana}-particles,
% in Boguliubov-deGennes descriptions.  \cite{read00,volovik}
%%
%%
Such zero-modes indicate a demarcation line 
that separates two topologically distinct phases
(e.g. condensate and trivial vacuum) from each other.  
The number of  zero modes relates, via the 
Atiyah-Singer index theorem\cite{aharonov79}, 
to topological charges of certain gauge field configurations 
that represent the holes as vortices in continuum descriptions.  

%%%%%%%%%%%%%%%%%%%%%%%%%%%%%%%%%%%%%%%%
%%%%%%%%%%%%%
A natural lattice representation of topological defects 
are vacancies in tight-binding models of graphene.
Indeed, an isolated vacancy carries a zero-energy mode 
that exhibits
the characteristic $1/(x+\ci y)$ decay 
away from the defect.\cite{PEREI08}
We can associate with every vortex 
a unit of topological charge, with a sign that is positive for 
one and negative for the other sublattice. Then, a first expectation 
based on the index theorem familiar from the continuum theory could be 
that a 
%number of zero modes is given by 
%$\Nzm=|\NA-\NB|$, where $\NA,\NB$ denote the number of carbon atoms 
%in the respective sublattices of the graphene sample. 
%Thus, a continuum description of graphene for a situation with a 
mismatch of vacancy concentrations in the two sublattices, $\cA,\cB$, 
would induce a density $\nzm=|{\cA{-}\cB}|$ of zero modes.
Moreover, in the limit of balanced sublattices, $\cA{=}\cB$, one 
would expect a cancellation of topological charges and therefore 
$\nzm{=}0$. 

%%%%%%%%%%%%%%%%%%%%%%%%%%%%%%%%%%%%%%
In this work, we are mostly interested in the spectra 
of tight-binding models of large (but finite) 
graphene flakes. In this case, a kind of lattice
analog of the continuum version of the Atiyah-Singer index theorem applies, 
\be
\label{e0}
\zeta \geq |\NA-\NB|,
\ee 
where $\NA,\NB$ denote the number of sites 
in the respective sublattices and $\zeta$ is the number of zero modes.
Eq. \eqref{e0}  has been derived for general bipartite lattices 
%models 
by Inue, Trugman and Abrahams and assumes a connected lattice graph.\cite{INU94} 
%{\bf FE: Check last statement.}
%%%
Since \eqref{e0} is an inequality, the index theorem can deliver only 
a lower bound and in the perfectly 
balanced situation, $\NA/\NB=1$, it does not give any information 
at all, strictly speaking. 
As we will demonstrate in the following, the tight-binding 
(tb-)model for graphene flakes exhibits a finite 
concentration of zero modes even in the situation of balanced sublattices.
A first discussion of possible consequences for 
graphene's material properties will be offered. 
%Indeed, he effect was observed first in the context of 
%molecular orbital theory of unsaturated hydrocarbons\cite{longuethiggins50,salemBook}. 
%Recently, a refined description in terms of benzoid graph theory has been achieved.
%\cite{fajtlowicz05}
%Our numerical work quantifies the exact qualitative statements in this work
%in the limit of macroscopic flake sizes.  

%%%%%%%%%%%%%%%%%%%%%%%%%%%%%%%%%%%%%%

%%%%%%%%%%%%%%%%%%%%%%%%%%%%%%%%%%%%%%%%%%%%%%%%%%%%%%%%%%%%%%%%%%%%%%%
%%%\section{Zero modes of isolated clusters: analytical results}
%% Further general remarks
%%%%%%%%%%%%%%%%%%%%%%%%%%%%%%%%%%%%%%%%%%%%%%%%%%%%%%%%%%%%%%%%%%%%%%5

A few preparatory remarks are in order. 
There are two reasons why the estimate based on 
 ``$=$'', as adopted in the continuum index theorem,  
tends to be a very crude one, when considering a 
graphene (tb-)lattice decorated with vacancies. 
%The first one is obvious. 
(i) At every non-zero concentration 
of vacancies, $\cA,\cB>0$, there is a non-vanishing probability that a sequence 
of vacancies punched into pristine graphene forms a closed line, i.e. a loop. 
This loop cuts the  sample into two separate fragments ("cluster"). 
The estimate Eq. \eqref{e0} applies to each fragment (indices 1,2) 
separately, $\Nzm\geq|\NA_1{-}\NB_1|+|\NA_2{-}\NB_2|$, $\Nzm{=}\zeta_1{+}\zeta_2$,
which, in general, gives a non-vanishing result for the total number of zero modes, 
$\Nzm$, per sample even if $|\NA_1{+}\NA_2{-}\NB_1{-}\NB_2|=0$. 
Generic continuum treatments do not incorporate  effects of 
correlated disorder and therefore fragmentation escapes their scope. 
%%%

(ii) Inue et al. were interested in localization properties 
of fermions on {\em connected} bipartite lattices (single cluster, no fragments) 
as representatives of the chiral universality class. \cite{INU94,EVE08} 
Their result is, in fact, a special case of a stronger inequality that has been 
proven earlier. Longuet-Higgins~\cite{longuethiggins50} showed in 1949, 
that the number of zero modes is bounded from below by  
$
\zeta \geq N - 2 \Ndb
$
where $\Ndb$ denotes the maximum number of double bonds,
that can be placed on a given graphene cluster (``flake'') 
with $N=\NA+\NB$ sites. 
\footnote{We here follow the 
nomenclature of the original literature.
The maximum number of placeable double-bonds, 
$\Ndb$, may also be thought of as the maximum number of 
non-adjacent edges, $\beta$. 
Alternatively, $\Ndb$ 
relates to the total number of sites, $N$, and the maximum number of 
non-adjacent sites, $\alpha$, via $\Ndb=N-\alpha$. 
}
This latter result, though derived for graphene flakes, is valid for the much more 
general situation of hopping on bipartite graphs. \cite{SuppInfo} Recent research
on benzoidal graph theory sharpens the statement further: 
on the honeycomb lattice the equality sign holds true, so that we have 
the exact statement\cite{fajtlowicz05} 
\be
\label{e0b}
\zeta = N - 2 \Ndb. 
\ee
As a consequence and as we further exemplify below, 
even a single balanced cluster, $\NA_i{=}\NB_i$, (cluster index $i$) 
exhibits, in general, a number of ``extra'' modes due to the mismatch $\Ndb < \NA,\NB$.
We call these modes ``supernumerary'', while the
conventional ones (due to 
sublattice imbalance) are called 
``predictable''. 
\footnote{We here follow the mathematical literature~\cite{rigby79}
deviating from the one used 
by Longuet-Higgins~\cite{longuethiggins50} 
and Bonfanti {\it et al.}~\cite{bonfanti11} 
who refer only to those modes as 
supernumerary that exist in excess of $N-2\Ndb$.
}
%On a qualitative level there is no difference between the two of them, 
%except that supernumerary modes appear in pairs, 
%while predictable do not. \cite{rigby79}
%%%%%%%%%%
Only in the simplest cases the supernumerary zero modes 
(\szm) can be traced back to 
particular structural elements,
such as singly-connected lattice sites
(coordination number one, ``dangling \ds{}s''). 
%such as dangling bonds.
In the present communication  we establish, using numerical simulations,  
that \szm{}  come with macroscopic abundance on large isolated clusters; 
in particular, their concentration on the percolation cluster is finite.  
%As it turns out, these modes dominate the DoS 
%for compensated disorder and large enough system sizes 
%at any non-vanishing vacancy concentration. 

%%%%%%%%%%%%%%%%%%%%%%%%%%%%%%%%%%%%%%%%%%%
%%%%%%%%%%%%%%%%%%%%%%%%%%%%%%%%%%%%%%%%%%%%%%%%%%%%%%
\begin{figure}[tb]
%\centering
\includegraphics[width=0.175\columnwidth]{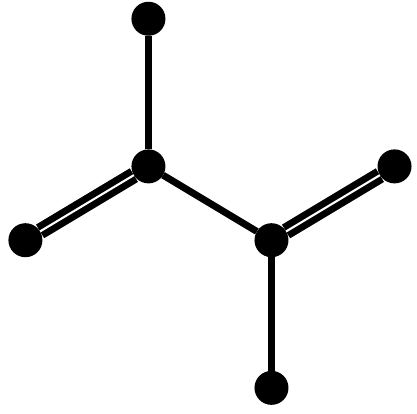}
\caption{The simplest example of a balanced lattice animal, $\NA{=}\NB{=}3$ 
that carries \szm. 
There are no predictable zero modes because $\NA=\NB$. 
However, two double bonds can be placed 
and therefore we have two \szm: $\zeta{=}6{-}4{=}{2}$.}
\label{f1}
\end{figure}
%%%%%%%%%%%%%%%%%%%%%%%%%%%%%%%%%%%%%%%%%%%%%%%%%%%%%%%%%%%%%%%%%%%%%%%%%%%%%%%%

%%%%%%%%%%%%%%%%%%%%%%%%%%%%%%%%%%%%%%%%%%%%%%%%%%%%%%%%%%%%%%%%%%%%%%%%%%%%%%%%
\subsection{Lattice animals, edge motifs and topological modes} 
%%%%%%%%%%%%%%%%%%%%%%%%%%%%%%%%%%%%%%%%%%%%%%%%%%%%%%%%%%%%%%%%%%%%%%%%%%%%%%%%
% {\it Lattice animals, edge motifs and topological modes.} 
To illustrate the efficiency of the equality Eq. \eqref{e0b} 
%we list in appendix \ref{a2} all clusters (``lattice animals'') 
we have investigated all clusters (``lattice animals'') 
up to 7 sites, i.e. mass $s{=}7$, and their associated spectra. 
In Fig.~\ref{f1} we show as an example the 
smallest animal that supports extra zero modes, $\zeta=2$, in this case. 
This result is verified from the associated connectivity matrix. 
\be
\label{e9}
h= 
\left(
\begin{array}{ccc}
 1 & &  \\
 1 \\
 1 & 1 & 1
\end{array}
\right).
\ee
It displays two identical pairs of columns, 
hence  $\zeta=2$

\begin{figure}[tb]
\centering 
\subfloat[]{\label{danglingb}
%\subfloat[A dangling bond]{\label{fig:danglingbond}
\includegraphics[width=0.22\columnwidth,keepaspectratio=true]{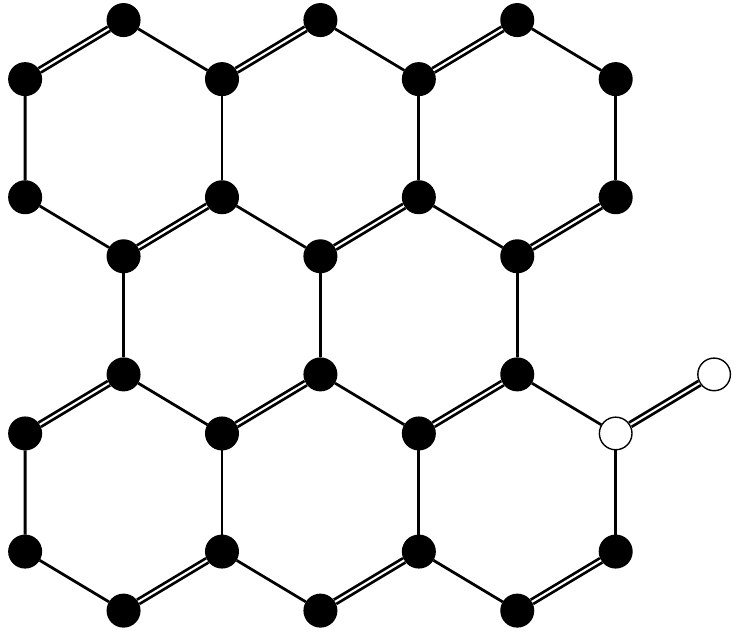}}
%\quad
\subfloat[]{\label{fig:ddbond}
\includegraphics[width=0.22\columnwidth,keepaspectratio=true]{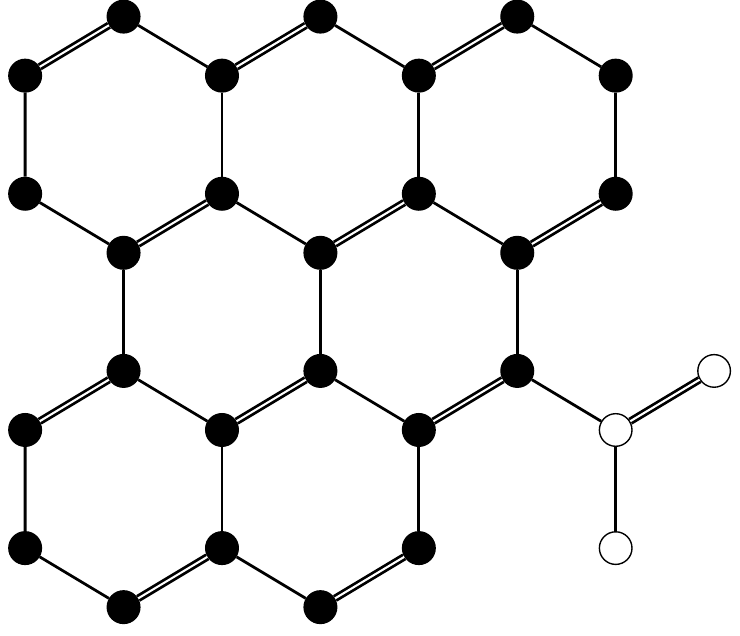}}
%\quad
\subfloat[]{\label{fig:dd13}
\includegraphics[width=0.24\columnwidth,keepaspectratio=true]{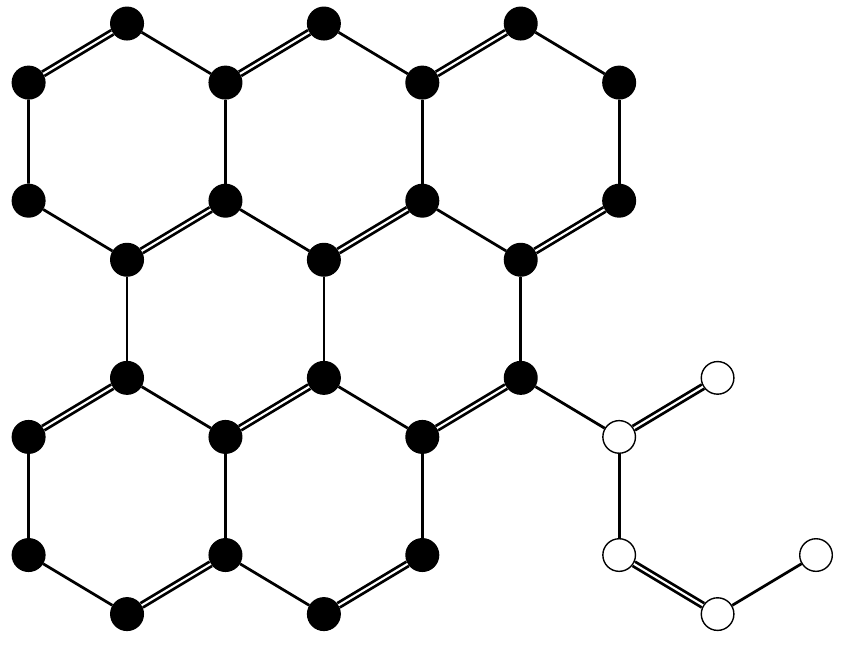}}
 % dd_13.eps: 0x0 pixel, 300dpi, 0.00x0.00 cm, bb=
%\quad
%\subfloat[$V$-structure]{\label{fig:V_structure}
\subfloat[]{\label{fig:U_struct}
\includegraphics[width=0.24\columnwidth,keepaspectratio=true]{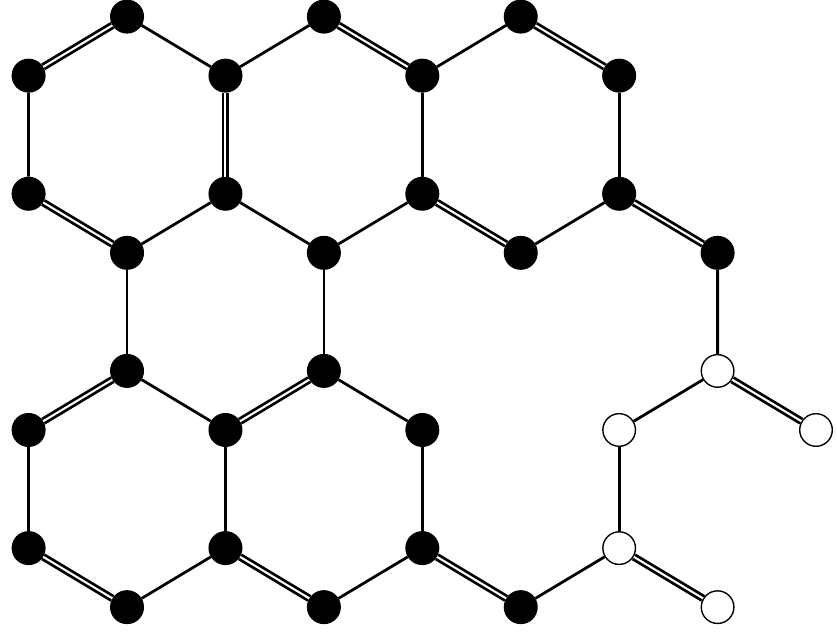}}
 % danglingbond.eps: 0x0 pixel, 300dpi, 0.00x0.00 cm, bb=14 14 363 474
 % ddbonds.eps: 0x0 pixel, 300dpi, 0.00x0.00 cm, bb=14 14 697 591
\caption[Dangling Bonds]
{Edge motifs: (a) dangling \ds. (b) double dangling \ds{} 
that is associated with a zero-energy state, as can be seen 
from the fact that one site without double bonds is identified.
(c) double-dangling chain. (d) ``$U$-structure''. }
\label{f2}
\end{figure}

The example of small animals illustrates that the proper placing 
of double bonds requires the knowledge of the topology of the full cluster. 
To find $\Ndb$ for very large clusters therefore is not a trivial task;
it is, in fact, exponentially hard. 
%%%
However, the structure of matrix \eqref{e9} suggests a recipe for generating 
structural elements of much larger clusters that carry a type of zero modes, 
that do not require scrutinizing the full object in order to be predicted. 
Such elements are shown in Fig.~\ref{f2}, 
dangling \ds{}s (left), double dangling \ds{}s (right). 
Double dangling \ds{}s constitute of two 
singly-connected sites, that share a common internal port-site. 
It is easy to see directly from placing double bonds 
that double dangling \ds{}s are associated with an extra zero mode. 
By inspecting the connectivity matrix the conclusion is confirmed 
immediately. 

%%%%%%%%%%%%%%%%%%%%%%%%%%%%%%%%%%%%%%%%%%%%%%%%%%%%%
%%%%%%%%%%%%%%%%%%%%%%%%%%%%%%%%%%%%%%%%%%%%%%%%%%%%%%
%%%%%%%%%%%%%%%%%%%%%%%%%%%%%%%%%%%%%%%%%%%%%%%%%%%%%%
Hence, one has a structure 
\be
H = 
\left(
\begin{array}{cc|ccc}
 1 	& 	1 & 	\star & \ldots & \star \\
 0 	& 	0 & 	\star & \ldots & \star \\
 \vdots & \vdots & \vdots & \ddots & \vdots \\
 0 	& 0 	& 	\star & \ldots & \star
\end{array}
\right)
\ee
The two linearly dependent column vectors reflect the occurrence of 
a zero-energy state. 
%{\tt Norman please check: I think we're going to have just one extra state, 
%do  you agree?}

Analogous arguments can be given for more complex edge structures involving motifs 
similar to double dangling \ds{}s, Fig.~\ref{f2}, right panels. 
For instance it can be shown  that all double dangling \ds{}s where both \textit{arms} consist of 
an odd number of sites lead to linearly dependent column vectors and consequently cause zero modes. 
%This is shown in Formula \eqref{dd13-bond Hamiltonian} for a double dangling bond with arms of 
%length $1$ and $3$, which is depicted in Figure \ref{f3}, left. 
%\be
%h=\left(\begin{array}{ccccccc}
%1&1&0&\star&\cdots&\star\\
%0&1&1&\star&\cdots&\star\\
%0&\star&0&\star&\cdots&\star\\\vspace{0.15cm}
%\vdots&\vdots&\vdots&\vdots&\vdots&\vdots\\
%0&\star&0&\star&\cdots&\star\\
%	\end{array} \right)
%	\label{dd13-bond Hamiltonian}
%\ee
Another edge motif associated with zero-modes, the `$U$-structure', 
is depicted in Fig.~\ref{f2}, right. It consists of a site of degree $2$ in the center 
which is connected to two sites adjacent to dangling \ds{}s. In this case 
the dangling \ds{}s can also be expanded to arms leading to a whole class of structures. 
As for the double dangling \ds{}s, structures having odd-numbered arms also produce zero modes.

%%%%%%%%%%%%%%%%%%%%%%%%%%%%%%%%%%%%%%%%%%%%%%%%%%%%%%%%%%%%%%%%%%%%%%%%%%%%%5555
%\begin{figure}
%\centering
%\subfloat[Double dangling carbons with arms of length $1$ and $3$]{\label{fig:dd13}
%\includegraphics[width=0.45\columnwidth,keepaspectratio=true]{pics/dd_13.eps}
% % dd_13.eps: 0x0 pixel, 300dpi, 0.00x0.00 cm, bb=
%\qquad
%\subfloat[$V$-structure]{\label{fig:V_structure}
%\includegraphics[width=0.45\columnwidth,keepaspectratio=true]{pics/M_structure.eps}
%\caption{Structural edge elements associated with zero energy states. Left: Double-dangling chains. Right: ``$U$-structure''.} 
%\label{f3}
%\end{figure}
%%%%%%%%%%%%%%%%%%%%%%%%%%%%%%%%%%%%%%%%%%%%%%%%%%%%%%%%%%%%%%%%%%%%%%%%%%%%%%%

%%%%%%%%%%%%%%%%%%%%%%%%%%%%%%%%%%%%%%%%%%%%%%%%%%%%%%%%%%%%%%%%%%%%%%%%%%%%%%%
The examples discussed thus far suggest the following picture: 
Eq.~\eqref{e0b} predicts zero modes of two different kinds. 
The number of predictable modes equals the sublattice mismatch $|\NA-\NB|$.
They are expected to be topological. 
There are an additional number of supernumerary 
zero modes that come in pairs and that exist 
even in the balanced case, $\NA=\NB$. 
%\footnote{The Stone-Wales defect is an example for a topological 
%mode in the balanced case. One easily convinces oneself that 
%%%Eq. \protect{\eqref{e0b}} 
%carbon placing predicts two zero modes.}

%%%%%%%%%%%%%%%%%%%%%%%%%%%%%%%%%%%%%%%%%%%%%%%%%%%%%%%%%%%%%%%%%%%%%%%%%%%%%%%%
\subsection{Percolation cluster}
%%%%%%%%%%%%%%%%%%%%%%%%%%%%%%%%%%%%%%%%%%%%%%%%%%%%%%%%%%%%%%%%%%%%%%%%%%%%%%%%
% {\it Percolation cluster.}
We discuss implications of our findings for the spectrum of a spanning cluster. 
In the limit of vanishing vacancy concentration,  $\nAB\to 0$, 
the percolation cluster is dense exhibiting mostly 
isolated vacancies; correlated formations of vacancies that would 
induce pairs of extra zero modes, e.g., due to edge motifs, are rare:  
\be
\label{e12}
%\zeta(s,n_\text{vac})=|\NA_\text{p}-\NB_\text{p}| + c \nAB^\alpha s + \text{subleading terms}
\zeta(s, n_\text{vac}) = \zp^{\text{pre}} + \zp^{\text{sup}}
\ee 
where $\zp^{\text{pre}} = | \NA_\text{p} - \NB_\text{p} |$ and $\zp^{\text{sup}} = \mathscr{F}[n_\text{vac}] s$, 
and $\mathscr{F}[n_\text{vac}] \propto n_\text{vac}^\alpha$ in the limit of vanishing concentration,  $n_{\text{vac}} \rightarrow 0$. 
Here the exponent $\alpha$ is a positive number and 
$s{=}\NA_\text{p}+\NB_\text{p}$ is the cluster mass. 
To estimate $\alpha$, we recall that the leading corrections  
are due to correlated disorder configurations. 
If the structures dominating the extra modes are given, e.g.,
by double-dangling \ds{}s 
we have a scaling with $n_\text{vac}^4$. Hence, one expects an inequality $\alpha \leq 4$. 

Now, consider the limit of very large clusters
%By very large we mean 
for which we expect a scaling 
$|\NA_\text{p}{-}\NB_\text{p}|\propto s^{1/2}$. 
%(for clusters drawn from an ensemble of balanced flakes) 
% reproducing the variance of 
% the binomial distribution. \cite{HUA10,EZA11}. 
% {\bf SB: Did I put this in??}
Therefore, the second term in Eq.~\eqref{e12} dominates for large enough systems: $s\gg \xi^2$. 
%Hence, the second term in Eq. (\ref{e12})
%dominates for large enough systems: . 
The crossover length $\xi$ exhibits a scaling with vacancy 
concentration: $\xi\propto  n_\text{vac}^{-\alpha}$. 

\begin{figure}[tb]
\centering
%\includegraphics[width=0.25\columnwidth,keepaspectratio=true]{Figure3_double_bonds}
%\hfill  
\includegraphics[width=1\columnwidth]{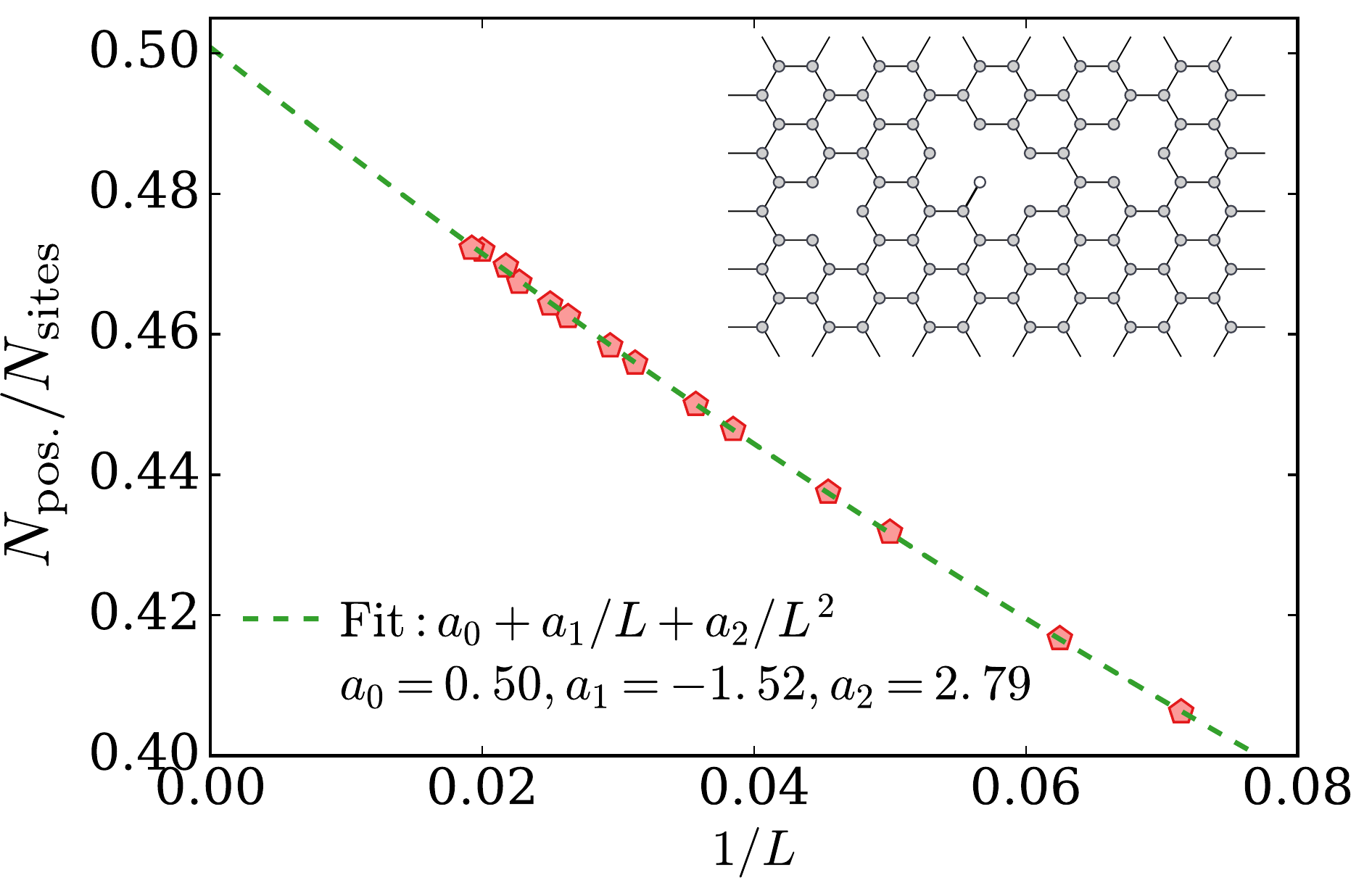}
\caption[Topological modes not predicted]
{Supernumerary zero modes (\szm) in balanced graphene sheets with a single dangling \ds{} and two isolated vacancies.
%Exact enumeration of all placements of the pair of isolated vacancies 
%that yield two \szm .
Plot shows that the number of possibilities, $N_\text{pos.}$, for all placements of the pair of isolated vacancies 
that yield two \szm, grows proportional to the sample area (number of lattice sites, $N_\text{sites}$).
Inset: Example with size $10\times 20$ that exhibits two \szm{}~with double-periodic boundary conditions, which is seen
by
either exact diagonalization or by bond placing.
\label{f3}}
\end{figure}

%%%%%%%%%%%%%%%%%%%%%%%%%%%%%%%%%%%%%%%%%%%%%%%%%%%%%
Motivated by the investigation of the example displayed in 
Fig.~\ref{f3} inset, 
we propose that $\alpha$ is smaller than four. 
The plot shows a balanced sheet with two vacancies per sublattice. 
One convinces oneself, e.g., by placing double bonds 
or by exact diagonalization 
that this example exhibits two supernumerary modes, $\zp^\text{sup}=2$. 
They are brought about by the combined effect of 
a dangling \ds{} and two single vacancies. 
Importantly, there is a large number of possibilities for 
placing the isolated vacancies with respect to the dangling \ds{} 
all yielding two supernumerary zero modes. 
For a given lattice size $L$ we have determined this number via exact 
enumeration assuming a toroidal geometry (double-periodic boundary conditions).
%%%
As seen from Fig.~\ref{f3} main, it is proportional to the sheet size 
as quantified, e.g., by the number of constituting lattice sites $L^2{-}4$.
Since this implies that getting two supernumerary modes out of four balanced 
vacancies simply is proportional to the probability of forming a dangling \ds, 
the example suggests $\alpha{=}2$. 
%%%

%%%%%%%%%%%%%%%%%%%%%%%%%%%%%%%%%%%%%%%%%%%%%%%%%%%%%%

We consider the analysis, Fig.~\ref{f3}, suggestive, but clearly it is 
far from conclusive, because  any argument employing 
a virial-type expansion is underlying an assumption of separability.
It is not obvious at this point to what extend such an assumption 
will hold for \szm. 

To illustrate the difficulty, we change the boundary conditions 
cutting the torus open, so it forms a zigzag-carbon nanotube. 
Then, two zero modes are found even in the absence of vacancies.
Under the assumption of perfect separability, one would expect that 
zigzag-nanotubes carrying a dangling \ds{} and two balancing 
vacancies should exhibit vacancy configurations with 
four zero modes, two boundary modes plus two extra modes related to the 
vacancies. This, however, is not the case; in all cases only two 
zero-modes are observed. Apparently, boundary modes tend to mix 
with the vacancy-induced modes and a naive notion of additivity
does not hold. To avoid boundary related 
complications, our focus has been on the torus geometry.

The experience with boundary-modes we interpret as a  
hint that separability is not guaranteed, 
as least not within the small sample sizes $L$ available to us.
This provides an additional motivation for us to proceed with the 
numerical analysis and investigate graphene sheets with 
a finite concentration of balanced vacancies.

%%

%%%%%%%%%%%%%%%%%%%%%%%%%%%%%%%%%%%%%%%%%%%%%%%%%%%%%%%%%%%%%%%%%%%%%%%%%%%%%%%%
%%%%%%%%%%%%%%%%%%%%%%%%%%%%%%%%%%%%%%%%%%%%%%%%%%%%%%%%%%%%%%%%%%%%%%%%%%%%%%%
\section{Zero-modes on percolation clusters: numerical simulation}
%%%%%%%%%%%%%%%%%%%%%%%%%%%%%%%%%%%%%%%%%%%%%%%%%%%%%%%%%%%%%%%%%%%%%%%%%%%%%%%
% {\it Zero-modes on percolation clusters: numerical simulation.}
We confirm our analytical considerations and demonstrate numerically 
that for compensated disorder the largest fraction of zero modes
is of the supernumerary kind, i.e. not predictable. 
Our  focus is on percolating clusters that we obtain in the
following way: 
A pristine flake of size $L\times L$ is generated with periodic boundary conditions. Vacancies are added at a given
concentration keeping $\NA=\NB$. 
The percolation cluster is the biggest fragment that remains from the original flake. 
We identify it  by employing the Hoshen-Kopelman algorithm (see Appendix for details).
%We relegate the description of methodological details to appendix \ref{app1}. 
%{\bf SB: fix reference appendix} 
% XXX
%%%%%%%%%%%%%%%%%%%%%%%%%%%%%%%%%%%%%%%%%%%%%%%
\begin{figure}[tb]
	\centering
\includegraphics[width=1.\columnwidth]{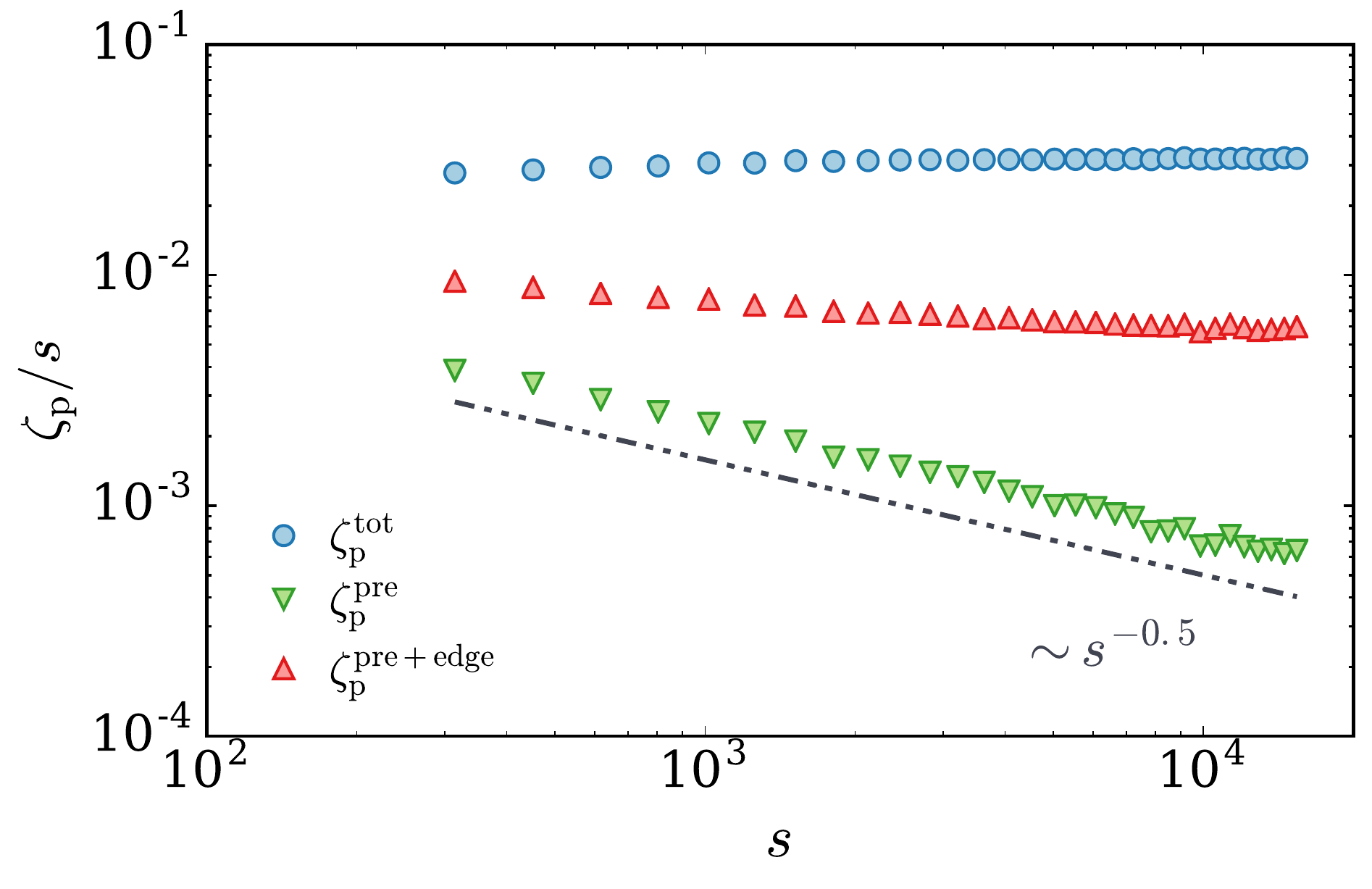}
 \caption[Scaling of the Number of Zero Modes on the Backbone with its Mass]{Average number of zero modes  associated 
with the percolation cluster obtained in systems of size $L{=}20,22,\ldots 128$ 
over its (average) mass $s$. The number of zero modes $\zp^{\text{pre}}$ based on sublattice imbalance and based on the 
combined effects of sublattice imbalance and edge structures are compared  
($\nAB{=}20\%$; average of $~1000$ realizations for systems up to $10000$ sites and over $100$ realizations for systems 
exceeding that size). In order to highlight the effect of edge motifs (Fig.~\ref{f3}) at large 
concentrations, we also show a curve ($\vartriangle$) where they have been added to the predictable modes. 
%{\tt what is the definition of red triangles: Is it really pre+edge or is it: all zero modes minus (pre plus edge)??
%Please check!} 
% {\tiny Norman, please dont show fit with exponent $m^{0.5025}$. Instead show a line as a guide to the 
% eye with $s^{1/2}$. 
% Do you think you could provide this plot with an inset that shows, how the saturation value for all 
% zero-modes depends on the vacancy concentration. Just those few points that you have will be enough. 
% They should be plotted over the concentration squared so as to exibit the correct power law.
% Moreover: Which edge-structures have been in included in $\Nzm^\text{imb. + edge}$? 
% Labeling: $\Nzm$ should be replaced with $\zp$.}
}
\label{f4}
\end{figure}
%%%%%%%%%%%%%%%%%%%%%%%%%%%%%%%%%%%%%%%%%%%%%%%

%%%%%%%%%%%%%%%%%%%%%%%%%%%%%%%%%%%%%%%%%%%%%%%%%%%%%%%%%%%%%%%%%%%%%%%%%%%%%%%%%%%%%%%%%%%%%%
\subsection{Results}
%%%%%%%%%%%%%%%%%%%%%%%%%%%%%%%%%%%%%%%%%%%%%%%%%%%%%%%%%%%%%%%%%%%%%%%%%%%%%%%%
% {\it Results.}
%% 
Fig.~\ref{f4} displays an ensemble averaged number of zero modes on the percolation cluster 
as a function of the cluster size at fixed vacancy concentration $\nAB=20\%$. 
We make three basic observations: 
(i) The expected scaling  $\sim \sqrt{s}$ for the predictable modes  is confirmed.
(ii) The total number of zero modes scales with $\sim s$, as expected. 
(iii) A considerable fraction of these modes is due to the simple (non-topological) 
edge motifs (dangling \ds{}s and `$U$'-type structures, see Fig.~\ref{f3}; for additional details see 
Ref.~\onlinecite{weik13}) that we discussed before. 
The probability of finding such structures scales with a 
power in $n_\text{vac}$ of order $4$ or higher. 
Due to our expectation $\alpha \lesssim 4$, 
we do not expect edge motifs, like Fig.~\ref{f2} (a,b,c) to 
give the dominant contribution in the low concentration regime. 
%This fraction grows with $s$, too, as one would expect. 
% We have performed a similar analysis also for vacancy concentrations $\nAB=5,10,15,20\%$. 
% In the inset of Fig.\ref{f4}
% we display the saturation value $\zeta(\nAB)$ in the limit of large system sizes. 
% It shows {\bf ... .  finish after inset has been produced}. 
%{\tt Norman, could you please provide that plot and check power laws also for edge-structures}

%%%%%%%%%%%%%%%%%%%%%%%%%%%%%%%%%%%%%%%%%%%%%%%
\begin{figure}[tb]
\centering
\includegraphics[width=1.\columnwidth]{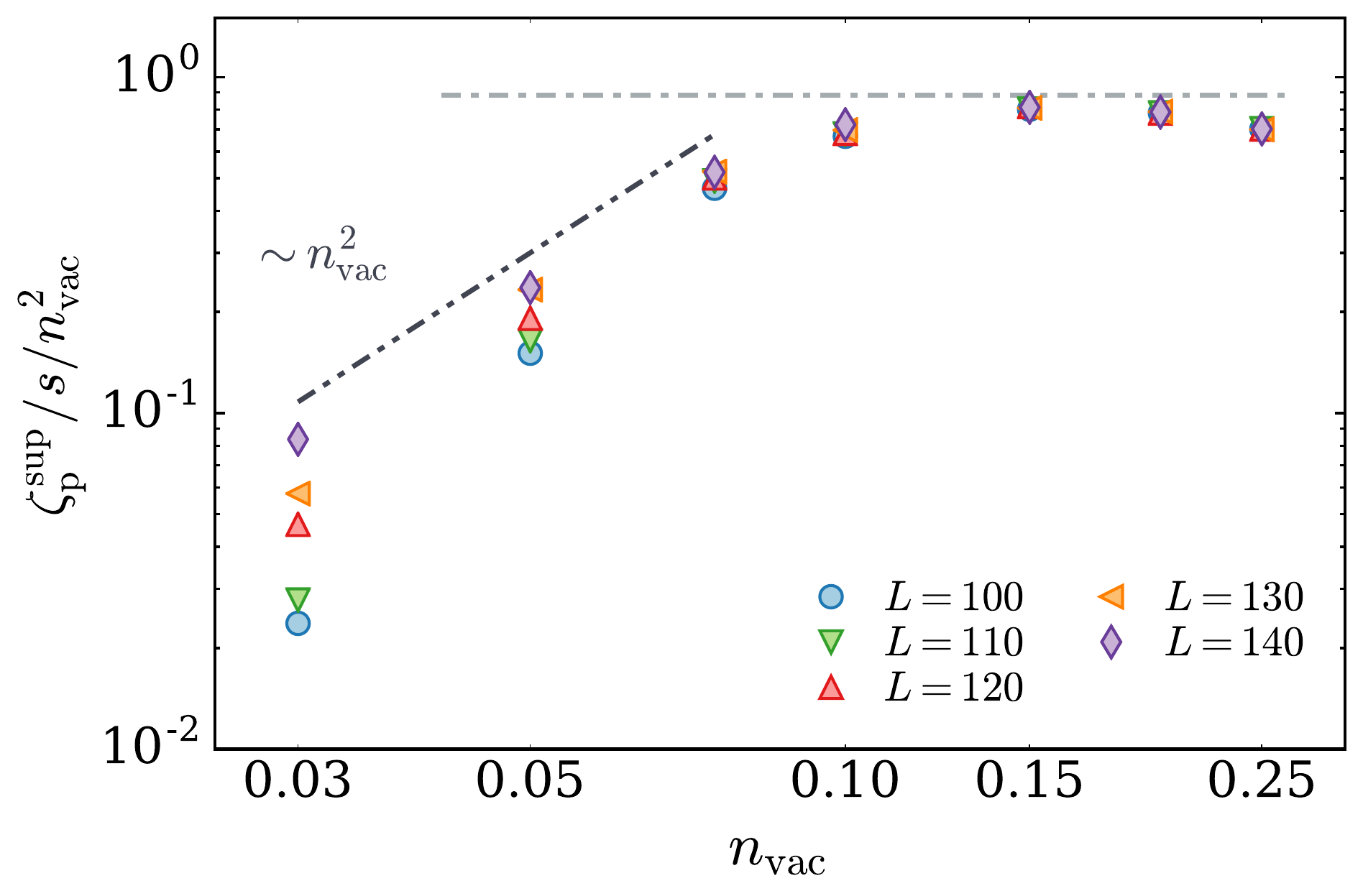}
 \caption[Scaling of the Number of supernumerary Zero Modes on the Backbone with its Mass]{Average number of 
supernumerary zero modes 
associated with the percolation cluster obtained in systems of size 
$L{=}100,110,\ldots 140$ over the vacancy concentration $n_\text{vac}$ in double logarithmic scale.  
The dashed-dot-dot line is a guide indicating a conservative upper bound~($\alpha=4$) for the power after finite size 
extrapolation. A value $\alpha=2$ corresponds to the horizontal line as shown in the figure~(dashed-dot-dashed). 
% {\tt Can we say something about error bars in the symbols?
% Shouldn't we give the horizontal line going through the data at 10\% ??}
}
  \label{f5}
\end{figure}
%%%%%%%%%%%%%%%%%%%%%%%%%%%%%%%%%%%%%%%%%%%%%%%
In Fig.~\ref{f5} we display how the number of supernumerary zero modes  
associated with the percolation cluster disappears with decreasing vacancy concentration. 
Percolation clusters have been drawn from systems of sizes $L=100, ..., 140$. 
At vacancy densities exceeding $\sim 7$\% the data can safely be extrapolated
into the macroscopic limit. However, at smaller concentrations finite size effects are 
substantial. As a consequence, the asymptotic limit of low $n_\text{vac}$, which would 
allow for a numerical determination of $\alpha$ is not readily observable.
Notice, however, that $\zp^{\text{sup}}$
increases by about a factor of three with $L$ changing by a factor of 
$\sim 1.4$ near vacancy concentration $\sim 3$\%. 
We would like to interpret this as an indication that $\zp^{\text{sup}}$ is 
overshooting the dot-dot-dashed line in Fig. \ref{f5} in the limit of 
large $L$, signalizing that indeed $\alpha<4$.
Hence, we believe that a relatively small value for the exponent
that would be consistent with the data at concentrations exceeding $\sim 10$\%, 
roughly $\alpha{\simeq} 2$, 
could conform with the existing numerical data also at lower vacancy concentrations.
%%%%%%%%%%%%%%%%%%%%%%%%%%%%%%%%%%%%%%%

\begin{figure}[tb]
 \centering
  \includegraphics[width=1.\columnwidth]{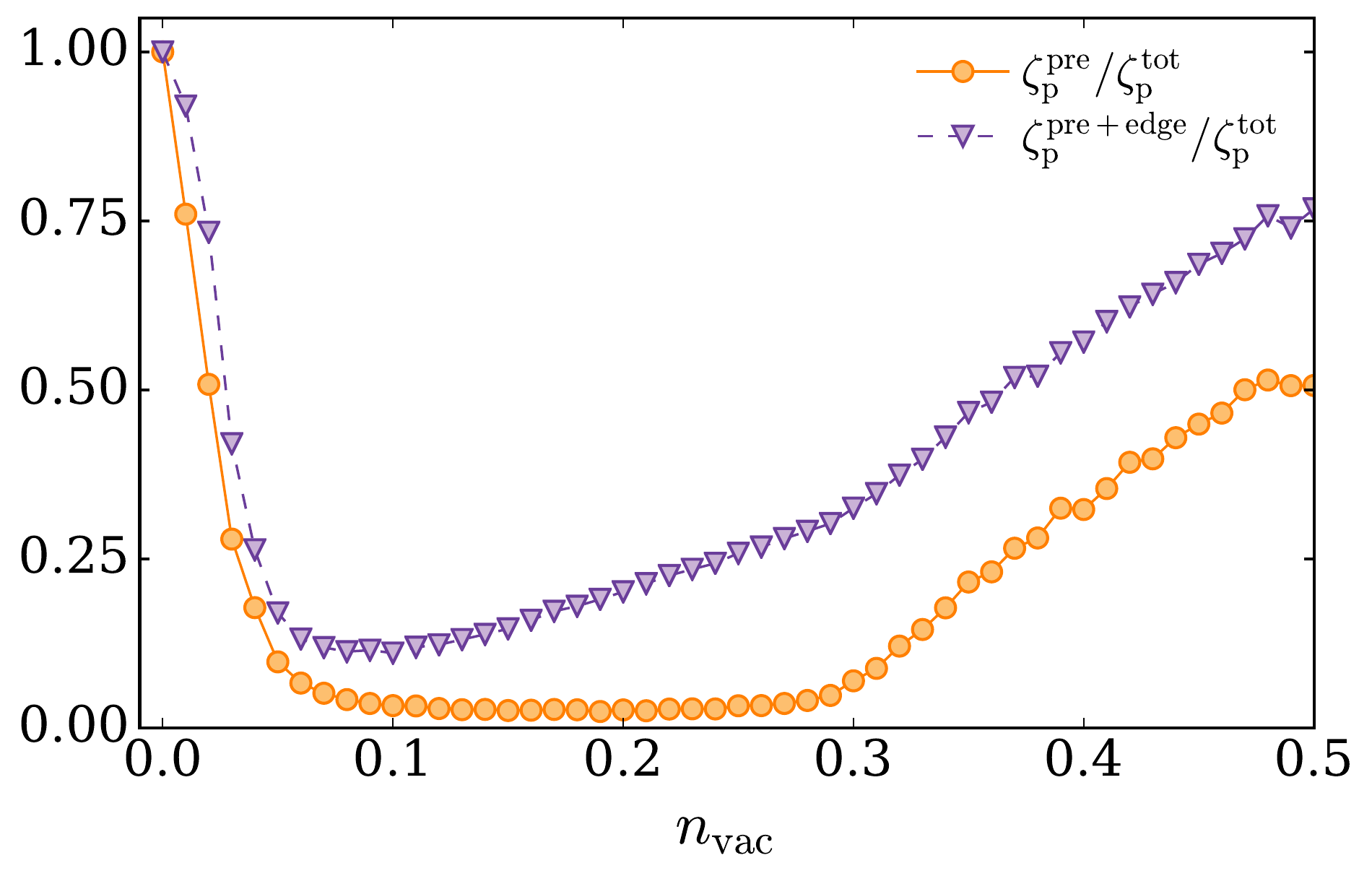}
% \includegraphics[width=1\columnwidth]{pics/Figxyz.eps}
 % zeromodes_backbone.eps: 0x0 pixel, 300dpi, 0.00x0.00 cm, bb=
 \caption[Zero Modes on the Backbone]
{Zero modes on percolation cluster. 
The total number of modes ($\triangledown$) and the fraction assigned to only sublattice imbalance 
($\circ$) is shown. Data is normalized to the total number of zero modes in the 
graphene sample (size: $L=100$; average over $500$ disorder realizations).
%{\tt What is shown here?? How can the total number of modes be the lowest lying curve??}
}
 \label{f6}
\end{figure}

Fig.~\ref{f6} shows once again an overview over a large range of 
vacancy concentrations. The fraction  
of zero modes that one finds on average on the percolation cluster is compared 
with the number of modes, $\zeta_\text{p}^\text{pre} = |\NA_\text{p}{-}\NB_\text{p}|$, predictable from the sublattice imbalance (at given system size, $L=100$). 
In the clean limit, $\nAB\to 0$, the crossover length exceeds our system size, 
$\xi\gg L$,  and all zero modes are seen to be predictable. 
%{\tt NW: why is it that the data in Fig. 4 does not approach unity 
%in the zero-concentration limit??}
%%% intermediate concentrations
Towards larger concentrations
the crossover length $\xi$ decreases and eventually in the range
$\nAB{\approx} 1\div3\%$ $\xi$ exceeds the system size $L=100$. 
In that regime, the supernumerary zero modes dominate 
and at concentrations above $7\%$ they contribute more than $80\%$. 
%{\tt What is the definition of the percolation cluster at vacancy concentrations 
%above the percolation threshold?}

%%%%%%%%%%%%%%%%%%%%%%%%%%%%%%%%%%%%%%%%%%%%%%%
%%%%%%%%%%%%%%%%%%%%%%%%%%%%%%%%%%%%%%%%%%%%%%%%%%%%%%%%%%%%%%%%%%%%%%%%%%%%%%%%%%%%%%%%%%%%%%

%%%%%%%%%%%%%%%%%%%%%%%%%%%%%%%%%%%%%%%%%%%%%%%%%%%%%%%%%%%%%%%%%%%%%%%%%%%%%%%
\subsection{A first discussion: Zero-modes and graphene's 
material properties}
%%%%%%%%%%%%%%%%%%%%%%%%%%%%%%%%%%%%%%%%%%%%%%%%%%%%%%%%%%%%%%%%%%%%%%%%%%%%%%%
Before the discussion of the impact of zero modes, 
we recall that vacancies in tight-binding lattices 
(tb-vacancies) are not meant to
model missing carbon atoms (real vacancies) 
in the graphene material.
For example, real vacancies give rise to strong lattice 
distortions and also, possibly, to edge reconstruction which is
not included in our tb-modeling. Instead, tb-vacancies are better 
thought of as a model of carbon atoms that have been promoted 
from $sp^2$- to $sp^3$-hybridization via chemical functionalization. 
% Wehling10, like in our prl from 2014
\cite{haefner14,wehling10} 

\subparagraph*{Unbalanced fragments.}
Despite being conceptually trivial,  the ``excess modes'' resulting 
from unbalanced fragments ($\NA_i{\neq}\NB_i$)  
can make an important contribution to the thermodynamic density of states (DoS). 
Therefore, they enter generic thermodynamic properties of the sample, and 
thus may be relevant, e.g, for magnetism. 
Indeed, the impact of zero modes for the electronic states 
and ground-state magnetism has been investigated intensively 
for small, isolated graphene-flakes with edge-hydrogenation.
\cite{yazyev10} 
For instance, a pair of supernumerary modes 
may give rise to an anti-ferromagnetic ground-state by virtue 
of Hund's rule coupling in a bow-tie shaped fragment called
``Clar's goblet''.\cite{wang09} To what extend this conclusion
carries over to larger systems is, however, not clear. 
Part of the difficulty is that supernumerary zero modes 
may not benefit from topological protection and therefore
can, in principle, be strongly susceptible to interaction
effects (beyond mean field).\cite{kinza10} 
We mention that electronic structure effects related to 
supernumerary modes have been reported 
to also manifest in preferred binding geometries. 
\cite{bonfanti11}

%%%%%%%%%%%%%%%%%%%%%%%%%%%%%%%%%%%%%%%%%%

\subparagraph*{Percolation cluster.}
The transport current is carried by the percolating (``spanning'')  
cluster; smaller fragments are irrelevant. Analogous to the 
thermodynamic DoS, also the transport DoS, which is 
associated with the percolation cluster, exhibits a singular behavior. 
% \be
% \label{e1} 
% \varrho(E) = \nzm \delta(E) + \text{smooth}.  
% \ee

Like any other fragment, the spanning cluster
is unbalanced, in general. Therefore, 
it supports a number of predictable zero modes, 
%$\zp^{\text{pre}}{\geq}|\NA_\text{p}{-}\NB_\text{p}|$, 
$\zp^{\text{pre}}{=}|\NA_\text{p}{-}\NB_\text{p}|$, 
even with (global) compensation $\NA{=}\NB$.
(`p' denotes the spanning/percolation cluster.)
%Second, the extra zero-modes contribute to $\zp$ as well.
%%%%%%%%%%%%%%%%%%%%%%%%%%%
Interestingly, our results indicate that the fraction of \szm{} 
contained in $\zp$ always dominates over the 
predictable modes at large enough cluster sizes, 
$\zp^\text{pre}\ll \zp^\text{sup}$. Therefore, 
unless \szm{} are typically localized they 
could leave a signature in transport 
calculations, which may not have been identified as of yet. 
%%%
In particular, it might be very difficult to resolve it in 
numerical studies relying, e.g, on the Kubo formula 
due to the smearing induced by the imaginary frequency shift. 
\cite{Mucciolo15} 
%%%%%%%%%%%%%%%%%%%%%%%%%%%%%%%%%%%

%%%%%%%%%%%%%%%%%%%%%%%%%%%%%%%%%%%%%%%%%%%%%%%%%%%%%%%%%%%%%%%%%%%%%%%%%%%%%%%
\section*{Conclusions}
%%%%%%%%%%%%%%%%%%%%%%%%%%%%%%%%%%%%%%%%%%%%%%%%%%%%%%%%%%%%%%%%%%%%%%%%%%%%%%%
% {\it Discussion.}
The implications of the existence of extra zero modes for our understanding of 
graphene lattice models with vacancy disorder have not previously been analyzed 
for macroscopic systems. The main finding of the present work is
that supernumerary zero modes exist in macroscopic abundance even in 
compensated graphene lattice models. 
These modes are not incorporated in present 
continuum theories of this material. Therefore, the question 
can be raised, what aspects of the hydrodynamic theory of  
bipartite lattice models generally coincide 
with the present continuum theories, as they have been studied, 
e.g., by Gade and Wegener\cite{gade91,gade93} and many others.
\cite{EVE08,RYU09,mirlin10,OST10a,KOE12,RYU12}

Our recent research has identified a low energy regime in which 
Gade-Wegner scaling of the DoS is violated. \cite{haefner14}
The similarity of the numerical data with concurrent field theoretical results 
was interpreted as evidence that local imbalances of the vacancy distribution 
produce very strong disorder effects that drive the Gade-Wegner 
fixed-point unstable. \cite{gornyi14, haefner14} Supernumerary zero modes, however, have not been considered in this work. 

One would expect, that 
the spectral properties away from zero energy should feel a certain impact of 
the (supernumerary) zero modes. We give two obvious reasons. First, the concentration of spectral weight 
at zero energy must lead to a depletion of spectral weight in other spectral regions. Second, eigenvectors at non-vanishing 
energies must be orthogonal to the (macroscopic) degenerate subspace at $E{=}0$. 
What this implies for physical observables, and whether a new energy regime at very low energies is 
introduced, remains to be seen. 
%%%%%%%%%%%%%%%%%%%%%%%%%%%%%55
\section*{Acknowledgements}
%%%%%%%%%%%%%%%%%%%%%%%%%%%%%55
%{\bf Acknowledgements.}
% \begin{acknowledgements}
Discussions with P. Ostrovsky, I. Gornyi, Ch. Stafford and 
R. Moessner are gratefully acknowledged. We are also grateful to V. H\"afner for his contribution at the early 
stages of this work.
SB thanks S. Roy for helping with the visualization. 
We thank I. Kondov and the J\"ulich Supercomputer Center 
(JUROPA, project HKA12) for computational assistance and resources. 
% \end{acknowledgements}
%%%%%%%%%%%%%%%%%%%%%%%%%%%%%%%%%%%%%%%%%%%%%%%%%%%%%%%%%%%%%%%%%%%%%%%%%%%%%%%%%

{\it Note added: }
After the completion of this manuscript we became aware of the study performed by  S. Sanyal, K. Damle and O. I.  Motrunich~\cite{Sanyal16} which also reports a finite density of zero modes in graphene with vacancies.

%%%%%%%%%%%%%%%%%%%%%%%%%%%%%%%%%%%%%%%%%%%%%%%%%%%%%%
\bibliography{lit}
%%%%%%%%%%%%%%%%%%%%%%%%%%%%%%%%%%%%%%%%%%%%%%%%%%%%%%
%%%%%%%%%%%%%%%%%%%%%%%%%%%%%%%% ##################################################################

%\end{document}
  
%\setcounter{equation}{0}
%\setcounter{figure}{0}
%\setcounter{table}{0}
%\setcounter{page}{1}
%\makeatletter
%\renewcommand{\theequation}{S\arabic{equation}}
%\renewcommand{\thefigure}{S\arabic{figure}}
%%\renewcommand{\bibnumfmt}[1]{[S#1]}
%%\renewcommand{\citenumfont}[1]{S#1}
\newpage
\appendix
% \section{Supplementary Information}

%%%%%%%%%%%%%%%%%%%%%%%%%%%%%%%%%%%%%%%%%%%%%%%%%%%%%%%%%%%%%%%%%%%%%%%
\section{Zero modes of isolated clusters: analytical results}
%%%%%%%%%%%%%%%%%%%%%%%%%%%%%%%%%%%%%%%%%%%%%%%%%%%%%%%%%%%%%%%%%%%%%%%
%{\it Zero modes of isolated clusters: analytical results.}
We consider the following representation
of all zero-energy modes in a graphene sample
\be
\Nzm = \sum_{\calC} \zeta(\calG_\calC)\  N_\calC(\cA,\cB)
\label{e2}
\ee
where the sum is over all types of clusters $\cal C$ 
(``lattice animals'') 
within the sample (including the percolating cluster) and 
$N_\calC$ denotes the number of clusters of a certain type. 
We have introduced here the number of zero modes 
associated with each cluster, $\zeta$, that is 
determined by the cluster's graph, $\calG$. The graph 
of a cluster is defined via the connectivity matrix, 
which may be thought of as the piece of the tight-binding 
Hamiltonian, $H_\calC$, associated with the cluster subspace.
All clusters sharing the same $H_\calC$ (up to rotations) 
are representatives of the same graph $\cal G$. 
%%%%%%%%%%%%%%%%%%%%%%%%%%%%%%%%%%%%%%%%%%%%%%%%%%%%%%%%%%%%%%%%%%%%

The cluster numbers $N_\calC$ and their dependency on the vacancy 
concentrations is a typical object of percolation theory. 
\cite{isichenko92}
It incorporates all statistical aspects of vacancy disorder
in \eqref{e2}  and, in particular, a classification of 
clusters according to their mass (total number of sites) and 
perimeter (number of sites forming the external boundary). 
\cite{STA94,isichenko92}
Quantum-mechanics enters \eqref{e2} via the weight-factor, $\zeta$, 
that is an intrinsic property of a given graph.  
%%%%%%%%%%%%%%%%%%%%%%%%%%%%%%%%%%%%%%%%%%%%%%%%%%%%%%%%%%%%%%%%%5

%%%%%%%%%%%%%%%%%%%%%%%%%%%%%%%%%%%%%%%%%%%%%%%%%%%%%%%%%%%%%%%%%5
\subsection{Inequalities and double bonds}
%%%%%%%%%%%%%%%%%%%%%%%%%%%%%%%%%%%%%%%%%%%%%%%%%%%%%%%%%%%%%%%%%5
%{\it Inequalities and double bonds.}
In order to reveal the connection between zero modes and placing of 
double bonds, we follow the original argument by Longuet-Higgins and reproduce his proof. 
To be specific, we consider a cluster of carbon atoms that form a graphene flake. 
Call the sublattice with the majority number of sites the A-sublattice, so $\NA>\NB$.
The goal is to show that 
\be
\label{e3}
\zeta \geq \NA+\NB-2\Ndb
\ee
for the number of zero modes associated with this cluster. 

{\em Proof:} The tight binding Hamiltonian, $H$, associated with the cluster 
can be represented as a matrix of the dimension $\dim H=\NA+\NB$ 
that takes the special block-off diagonal form
\be
H =
\left( 
\begin{array}{cc}
 0 & h \\
h^\dagger & 0 
\end{array}
\right) 
\ee
where $h$ is a $\NA\times \NB$-matrix that connects 
the A/B-sublattices with each other. We start by recalling that
\be
\zeta = \dim H - \text{rank } H . 
\ee  
The $\text{rank }$ of $H$ can be defined in several equivalent ways. 
One definition is to say that $\text{rank } H$ is given by the 
largest order of any non-zero minor of $H$. The use of this definition will 
in the end connect the proof to the concept of double-bonds. 

Namely, consider any term that arises when calculating 
a minor of $H$; it has the structure: 
\be
\label{ea}
\prod^{P} \bar h_{\bar a\bar b} \prod^{Q} h_{ab}. 
\ee
(For notational simplicity, we define here the matrix elements 
$[h^\dagger]_{ab}=\bar h_{ab}$, $a=1,\ldots, \NA, b=1,\ldots,\NB$.) 
To visualize this term, we can imagine the set of lattice points $\{a\},\{b\}$ that represent 
the sublattice A and B respectively. Let $\bar h_{\bar a\bar b}$ denote a double bond that 
connects the pair of sites $(\bar a,\bar b)$. Similarly $h_{ab}$ connects the pair 
of lattice points $(a,b)$ via a single bond. Importantly, since the minor is a determinant 
(of a submatrix of $H$) it is guaranteed that there cannot be a site which attaches to more 
than one double bond or to more than one single bond. 

Here, we have attributed a covering of the lattice sites with a bond pattern that follows 
a simple rule. We can also work the other way round and convert a lattice covering with 
single and double bonds, that follows these rules, into a term that appears when evaluating 
minors of $H$. This is what we will do, now. The goal is to find a covering such that the number
of factors appearing in Eq. \eqref{ea}, $P+Q$, becomes as large as possible. This largest 
possible value coincides with $\text{rank }H$.
 
The covering we are after is identified as follows: We cover the set of points ${a},{b}$ with 
the maximum number of double bonds that we can place, so $P=\Ndb$. The single bonds we place along 
the same pairs, $Q{=}\Ndb$. Therefore $\text{rank }H \leq2\Ndb$ and the proof is complete. 
Note, that the proof holds for any bipartite lattice, not only for the honeycomb one. 
%%%FE: Add a remark that explains, why leq and not just equal. 

\section{Numerical Method}
\label{app1}
 
The quantitative analysis combines two numerical approaches. On the one hand, a brute force diagonalization of the TB-Hamiltonian with vacancies based on Lapack's SVD routine is performed. On the other hand the microscopic structure of the sample is investigated by performing a cluster analysis based on a modified Hoshen-Kopelman cluster labeling algorithm\cite{HOS76}. By combining the two approaches the spectral properties of single clusters and microscopic structures on these clusters can be analyzed. Our present 
analysis is limited to $\mathcal{O}(10^4)$ sites due to the vanishing level spacing making the identification of exact zero modes challenging in larger systems.

%%%%%%%%%%%%%%%%%%%%%%%%%%%%%%%%%%%%%%%%%%%%
\subsection{The Hoshen-Kopelman algorithm}
%%%%%%%%%%%%%%%%%%%%%%%%%%%%%%%%%%%%%%%%%%%%

For the microscopic analysis of the generation of zero modes, a modified Hoshen-Kopelman cluster labeling algorithm for honeycomb structures \cite{HOS76} is employed. It allows to identify and label all clusters and to determine their sublattice imbalance. 

\begin{figure}[tb]
	\centering
		\includegraphics[width=0.42\columnwidth]{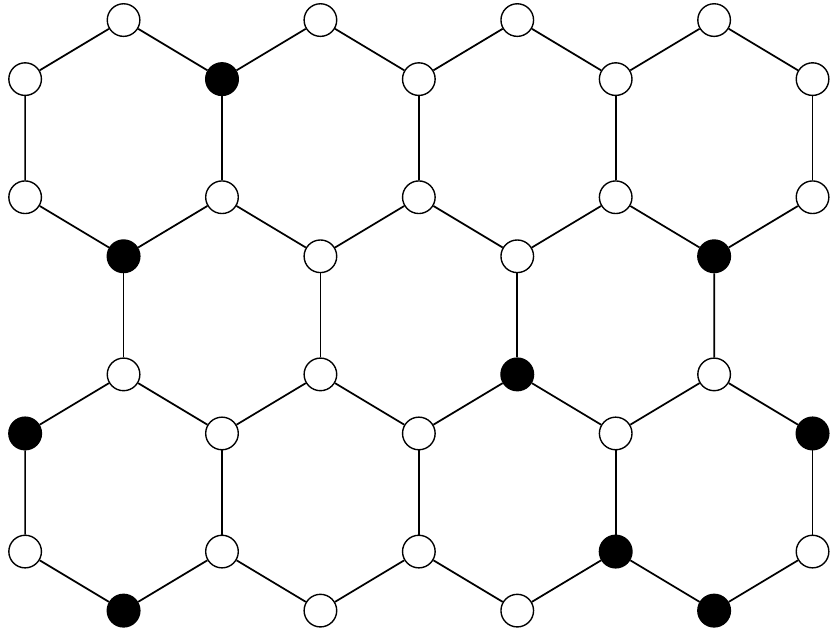}\qquad
		\includegraphics[width=0.42\columnwidth]{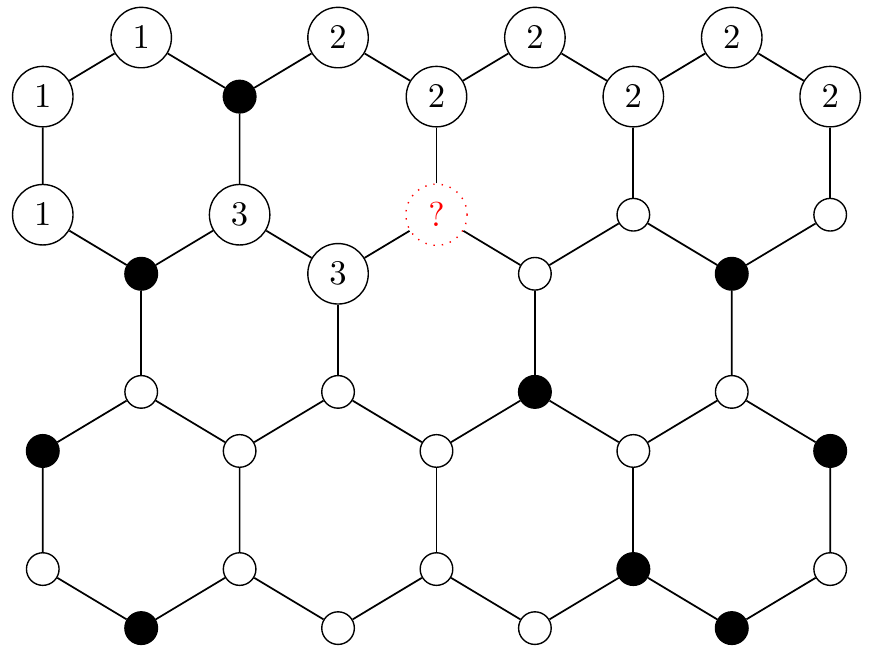}		
  \caption{Cluster labeling. Left: A exemplary graphene sample with vacancies (black). Right: Schematics of the Hoshen-Kopelman algorithm. The sites are labeled line wise: A given site is attributed the label of the sites to the left or the line above if one of them is non-vacant or they both have the same label. The marked site points out a conflict where the label to the left and to the top do not match and the clusters with label $2$ and $3$ get merged.}
  	\label{fig:hk}
\end{figure}

The lattice algorithm  scans line wise for non-vacant sites from the upper left to the lower right corner. (See Figure \ref{fig:hk}, left.) If a given non-vacant site is connected to an already labeled site it is given the same label. It is given a new label in case the adjacent sites  known so far are vacancies. However, a difficulty arises if the site is adjacent to two labeled sites having different labels (see Fig.~\ref{fig:hk} right). In this case two formerly independent clusters are merged requiring backward relabeling of sites, which is numerically very expensive. This problem was a major concern in the early days of percolation theory in the 50es and it was common belief that "the direct simulation of percolation is out of the question" \cite{HAM55} at that time.
%%%%%%%%%%%%%%%%%%%%%%%%%%%%%%%%%%%%%%%%%%%%
%%%%%%%%%%%%%%%%%%%%%%%%%%%%%%%%%%%%%%%%%%%%

%%%%%%%%%%%%%%%%%%%%%%%%%%%%%%%%%%%%%%%%%%%%
\begin{figure}[tb]
	\centering
		\includegraphics[width=0.6\columnwidth]{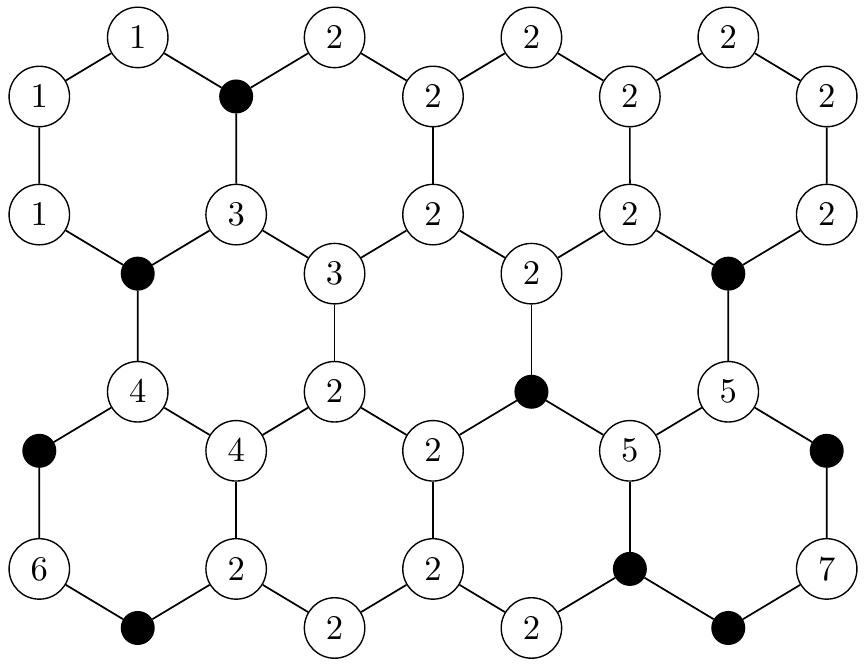}		
  \caption{The final labeling after the Hoshen-Kopelman algorithm has been performed.}
  	\label{fig:hk2}
\end{figure}
%%%%%%%%%%%%%%%%%%%%%%%%%%%%%%%%%%%%%%%%%%%%

The problem was solved in the middle of the 1970s with the development of the very efficient Hoshen-Kopelman algorithm \cite{HOS76} which allows to label all clusters with a single sweep through the lattice. It avoids backward labeling by introducing an additional array of cluster \textit{markers} which 
\begin{enumerate}
\item counts the number of sites belonging to a given cluster and
\item keeps track of the relabeling of clusters which have been merged with other (larger) clusters by storing a pointer to the larger cluster it has gotten merged with. In particular, the labels of clusters which have become merged with larger ones are not attributed to new sites any more (see Figure \ref{fig:hk2}). The sample may be cleaned by adding a second sweep through the lattice and replacing "antiquated labels" (like $3$ and $4$ in Fig. \ref{fig:hk2}). However, it is not required since all the information is stored in the Marker array. 
\end{enumerate}
For more detail on the Hoshen-Kopelman algorithm the reader is referred to the original paper by Hoshen and Kopelman \cite{HOS76}.

%%%%%%%%%%%%%%%%%%%%%%%%%%%%%%%%%%%%%%%%%%%%
\subsection{Modifications to the Hoshen-Kopelman algorithm}
%%%%%%%%%%%%%%%%%%%%%%%%%%%%%%%%%%%%%%%%%%%%

In this work some minor modifications were made to the Hoshen-Kopelman algorithm. 

\begin{itemize}
\item Periodic boundary conditions (PBC) are used instead of fixed ones in order to avoid zero modes due to zig-zag edges already discussed in the context of carbon nanoribbons \cite{NAK96}. They are integrated by adding an additional sweep along the sample's left and bottom edge once all sites have been labeled.
\item In addition to the storage of each cluster's number of sites in the marker array, its number of carbon atoms in the two sublattice are also kept s.t.~the sublattice imbalance of each cluster can be determined.
\item For investigating the properties of specific clusters such as the largest (spanning) one a second scan of the sample is added removing all but the desired cluster(s) from the sample. This clusters spectrum can then be determined by treating all other sites as vacant and diagonalizing the resulting TB-Hamiltonian. Hence we can directly determine the spectral properties of single clusters in the sample.
\end{itemize}

%%%%%%%%%%%%%%%%%%%%%%%%%%%%%%%%%%%%%%%%%%%%
\subsection{Identification of edge motifs}
%%%%%%%%%%%%%%%%%%%%%%%%%%%%%%%%%%%%%%%%%%%%

For the identification of edge motifs generating zero modes such as double dangling \ds{}s an additional analysis is
required once all clusters have been identified using the Hoshen-Kopelman algorithm. 
Our analysis of edge structures includes all structures up to double dangling bonds with one arm of length $1$ and one
of length $3$ and $U$-structures with arms of length $1$ (see Fig. \ref{f2}). These structures all involve
(coordination one) corner sites that are connected 
to only one other lattice site.

%%%%%%%%%%%%%%%%%%%%%%%%%%%%%%%%%%%%%%%%%%%%
\begin{figure}[b]
%\centering 
%\subfloat[A dangling bond]{\label{fig:danglingbond}
%\subfloat[A double dangling bond with arms of length one]{\label{fig:DDbond}
\includegraphics[width=0.27\columnwidth,keepaspectratio=true]{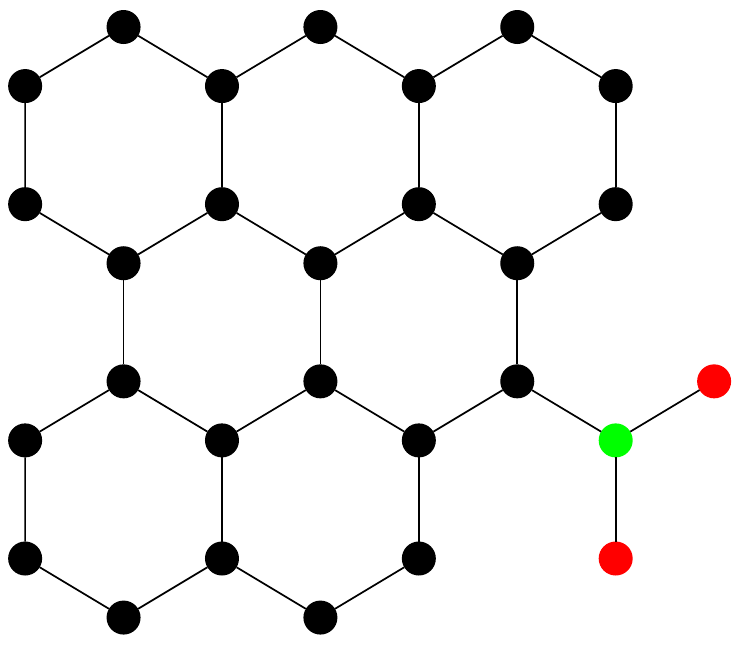}\hfill
%\quad
\includegraphics[width=0.3\columnwidth,keepaspectratio=true]{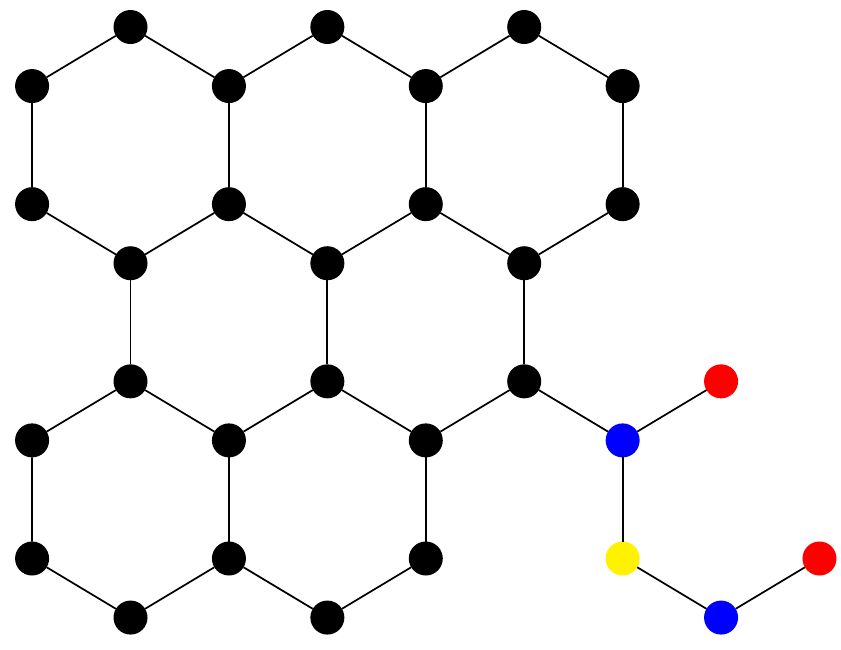}\hfill
 % dd_13.eps: 0x0 pixel, 300dpi, 0.00x0.00 cm, bb=
%\quad
%\subfloat[$V$-structure]{\label{fig:V_structure}
\includegraphics[width=0.3\columnwidth,keepaspectratio=true]{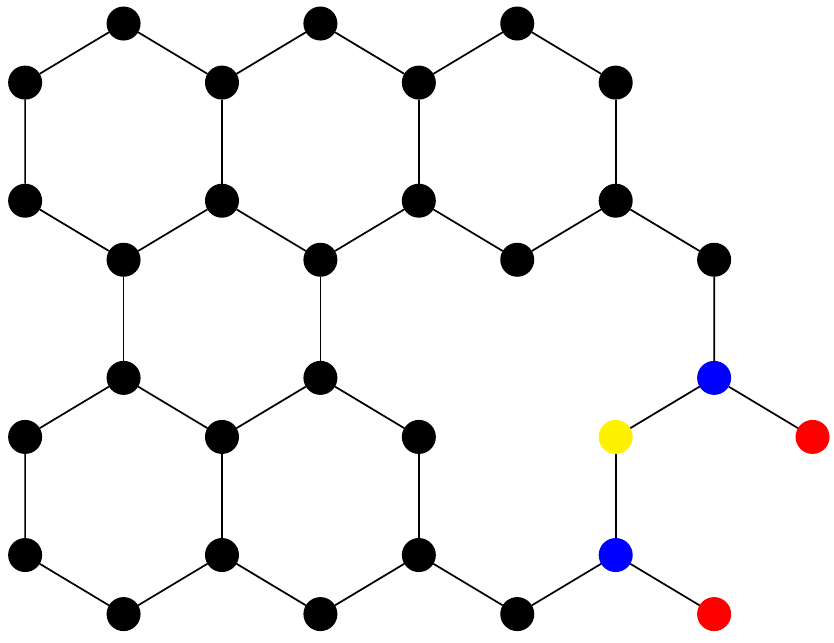}
\caption{Illustration of the algorithm for the identification of edge motifs using a coloring scheme. First step (red): Identification of corner sites. Second step (Blue):Coloring of sites adjacent to corner sites. In case an already blue site is to be colored this site is colored green. Third step (yellow): Coloring of sites which are adjacent to one vacancy and two blue (or green) sites. }
\label{fig:coloring}
\end{figure}
%%%%%%%%%%%%%%%%%%%%%%%%%%%%%%%%%%%%%%%%%%%%

Starting with these sites, the algorithm consists of three steps: 
\begin{itemize}
\item All corner sites are colored red.
\item The sites connected to red sites are colored blue. In case a site has already been colored blue, it is colored green. 
\item Sites adjacent to two blue- (or green-)colored sites and a vacancy are colored yellow. 
\end{itemize}

Obviously, the green sites mark the position of double dangling \ds{}s with arms of length $1$ since they are adjacent
to two (red) corner sites. The yellow sites mark the positions of double dangling \ds{}s with arms of length $1$ and
$3$ or $U$-structures (see Fig. \ref{fig:coloring}). 
%Hence, in combination with the Hoshen-Kopelman algorithm, the edge motives introduced in %Fig. \ref{f2} can be accounted for on each cluster. 
\end{document}